\documentclass[aps,prb,twocolumn,amsmath,amssymb,superscriptaddress,showpacs]{revtex4}
\usepackage{graphicx}
\usepackage{dcolumn}
\usepackage{color}
\usepackage{bm}

\begin{document}

\title{The quasiparticle band gap in the topological insulator Bi$_2$Te$_3$}

\author{I.~A. Nechaev}
 \affiliation{Donostia International Physics Center (DIPC), 20018 San Sebasti\'an/Donostia, Basque Country, Spain\\}
 \affiliation{Tomsk State University, 634050, Tomsk, Russia\\}

\author{E.~V. Chulkov}
 \affiliation{Donostia International Physics Center (DIPC), 20018 San Sebasti\'an/Donostia, Basque Country, Spain\\}
 \affiliation{Tomsk State University, 634050, Tomsk, Russia\\}
 \affiliation{Departamento de F\'{\i}sica de Materiales UPV/EHU, Facultad de Ciencias Qu\'{\i}micas, UPV/EHU, Apdo. 1072, 20080 San Sebasti\'an/Donostia, Basque Country, Spain\\}
 \affiliation{Centro de F\'{\i}sica de Materiales CFM - MPC, Centro Mixto CSIC-UPV/EHU, 20080 San Sebasti\'an/Donostia, Basque Country, Spain\\}

\date{\today}

\begin{abstract}
We present a theoretical study of dispersion of states which form the bulk band-gap edges in the three-dimensional topological insulator Bi$_2$Te$_3$. Within density functional theory, we analyze the effect of atomic positions varying within the error range of the available experimental data and approximation chosen for the exchange-correlation functional on the bulk band gap and \textbf{k}-space location of valence- and conduction-band extrema. For each set of the positions with different exchange-correlation functionals, we show how many-body corrections calculated within a one-shot $GW$ approach affect the mentioned characteristics of electronic structure of Bi$_2$Te$_3$. We thus also illustrate to what degree the one-shot $GW$ results are sensitive to the reference one-particle band structure in the case of bismuth telluride. We found that for this topological insulator the $GW$ corrections enlarge the fundamental band gap and for certain atomic positions and reference band structure bring its value in close agreement with experiment.
\end{abstract}


\maketitle

\section{\label{sec:introduction}Introduction}

Recently, it has been shown that bismuth telluride (Bi$_2$Te$_3$) is a topological insulator with a non-degenerate surface state\cite{Zhang_NatPhys_2009,Eremeev_JETPLett_2010,Eremeev_Nat_Comm_2012} forming a Dirac cone at the $\bar{\Gamma}$ point in a bulk band gap. The appearance of such a surface state is caused by the spin-orbit-induced inversion of the bulk band-gap edges. Low-energy quasiparticles in a two-dimensional electron system formed by the surface-state electrons behave as massless spin-helical Dirac fermions.\cite{Hsieh_Nature_2009} Properties of these quasiparticles depend ultimately on the dispersion of the bulk valence and conduction bands which shape the bulk band-gap edges, especially, within the band-inversion region. This fact revives the interest in a proper description of the bulk band structure of bismuth telluride.\cite{Kioupakis_PRB_2010,Yazyev_PRB_R_2012}

\begin{figure}[tbp]
\centering
 \includegraphics[angle=0,scale=0.7]{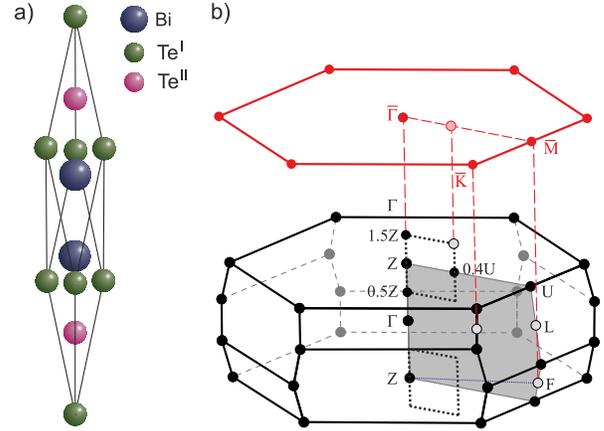}
\caption{(Color online) a) Rhombohedral unit cell of bismuth telluride. b) The bulk rhombohedral (at the bottom) and corresponding two-dimensional (at the top) Brillouin zones. The shaded area marks the high-symmetry mirror plane. The dotted-line rectangle outlines the $k$-space plane, where the dispersion of the uppermost valence band and the lowest conduction band is treated. Here Z=(0.5, 0.5, 0.5) and U=(0.823, 0.3385, 0.3385) as presented in reciprocal-lattice-vector coordinates.\cite{remark_coords}}\label{fig1}
\end{figure}

Being also a narrow gap semiconductor with properties promising for thermoelectric applications,\cite{Termo_lit} bismuth telluride has a quite long history of experimental and theoretical band-structure investigations. Density functional theory (DFT) calculations performed during the past few decades\cite{Oleshko_FTT_1985, Pecheur_PLA_1989, Thomas_PRB_1992, Mishra_JPCM_1997, Larson_PRB_2000} have revealed noticeable discrepancy between theoretical (50-130 meV) and experimental\cite{Austin_PPS_1958, Li_JAP_1961, Sehr_JPChSol_1962, Thomas_PRB_1992} (150-220 meV) data on the bulk band gap. From these calculations, it follows that Bi$_2$Te$_3$ has an indirect band gap with the conduction-band minimum (CBM) that occurs along the $\Gamma-$Z line and, therefore, has the multiplicity $M=2$, which disagrees with the experimental finding of $M=6$.\cite{Koehler_PSS_1976_2} As to the valence-band maximum (VBM), this extremum is located on the Z$-$F line that belongs to the mirror $yz$-plane of the Brillouin zone (BZ) (see Fig.~\ref{fig1}), which ensures the multiplicity $M=6$ in agreement with the experimental observation.\cite{Koehler_PSS_1976_1}

%=============================================================================================================
%=============================================================================================================

\begin{figure*}[tbp]
\centering
 \includegraphics[angle=0,scale=0.75]{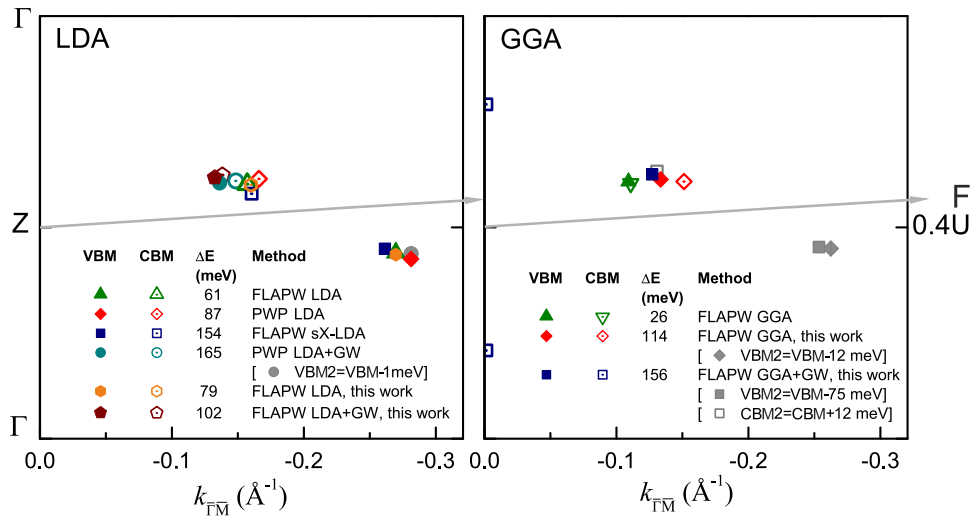}
\caption{(Color online) VBM and CBM locations as obtained with the use of different methods (FLAPW LDA from
Ref.~\onlinecite{Youn_Freeman_PRB_2001}, PWP LDA and PWP LDA+$GW$ from Ref.~\onlinecite{Kioupakis_PRB_2010},
FLAPW sX-LDA from Ref.~\onlinecite{Kim_Freeman_PRB_2005}, FLAPW GGA from
Ref.~\onlinecite{Wang_Cagin_PRB_2007}). Results of the present work correspond to the relaxed atomic positions of the set ``II'' (see the text).}\label{fig2}
\end{figure*}

In the last decade, the authors of Ref.~\onlinecite{Youn_Freeman_PRB_2001} have made the first attempt to answer the question of whether the CBM and VBM locations reported early to be on the $\Gamma-$Z and Z$-$F lines are true extrema. With the use of the full-potential linearized augmented plane-wave (FLAPW) method\cite{FLAPW} within the local density approximation (LDA) for the exchange-correlation (XC) functional it has been shown that the band extrema locate off the mentioned lines but in the mirror plane (see Fig.~\ref{fig2}), i.e., both extrema have $M=6$. The calculations have been performed for a rhombohedral crystal structure (see Fig.~\ref{fig1}) with experimental lattice parameters and atomic positions taken from Ref.~\onlinecite{Nakajima_JPCS_1963}. The authors of Ref.~\onlinecite{Youn_Freeman_PRB_2001} have found the VBM (hereafter referred to as the extremum X) at \textbf{k}=(0.546, 0.383, 0,383) and the CBM at \textbf{k}=(0.663, 0.568, 0.568) as presented in reciprocal-lattice-vector coordinates. The resulting ``fundamental'' band gap was obtained to be of 61~meV, which is even farther from the aforementioned experimental data than it followed from the previous calculations. Additionally, a second-highest VBM (hereafter referred to as the extremum C) that is 3.8~meV lower than the first one and a second-lowest CBM (on the $\Gamma-$Z line), which is about 50 meV higher than the CBM, have been observed at \textbf{k}=(0.665, 0.586, 0.586) and \textbf{k}=(0.273, 0.273, 0.273), respectively.

%=============================================================================================================
%=============================================================================================================

In order to improve the theoretical result on the band gap, in Ref.~\onlinecite{Kim_Freeman_PRB_2005} the calculation method of Ref.~\onlinecite{Youn_Freeman_PRB_2001} with the screened-exchange LDA (sX-LDA) aproach\cite{Bylander_Kleinman_PRB_1990} instead of the conventional LDA has been used. The calculations have revealed that the locations of the VBM and the CBM are slightly changed [(0.555,0.397,0.397) and (0.646,0.549,0.549), respectively], while the band gap runs up to 154~meV, which significantly improves the agreement with the experiment. These results give an insight to what extent the location of the band extrema and the band gap can be sensitive to an approximation to the XC functional. Additionally, in Ref.~\onlinecite{Kim_Freeman_PRB_2005} for the first time the effective mass parameters for the holes and the electrons in the vicinity of these ``true'' (in the sense of the DFT band structure explored over the whole mirror plane) extrema have been calculated. As compared with the available experimental data\cite{Koehler_PSS_1976_1, Koehler_PSS_1976_2} and with the effective-mass parameters calculated for the VBM and the CBM found along $\Gamma-$Z and Z$-$F lines, the parameters of Ref.~\onlinecite{Kim_Freeman_PRB_2005} demonstrate impressive improvement.

An investigation aimed at revealing the effect of crystal-structure relaxation on the extrema locations and the energy band gap has been performed in Ref.~\onlinecite{Wang_Cagin_PRB_2007} with the use of the FLAPW method as implemented in the FLEUR code\cite{FLEUR} within the generalized gradient approximation (GGA) of Ref.~\onlinecite{GGA_PBE} for the XC functional. Having started with experimental lattice parameters and atomic positions taken from Ref.~\onlinecite{Wyckoff}, after relaxation the authors of Ref.~\onlinecite{Wang_Cagin_PRB_2007} have ended up with a larger unit cell. As expressed in real-lattice-vector coordinates, the atomic positions have remained practically unchanged except for the position of Te$^{\mathrm{II}}$ atoms (see Fig.~\ref{fig1}). The band gap has grown from the unrelaxed value of 26~meV to 49~meV. The fact that such a structure optimization has practically doubled the band gap (as well as the use of the sX-LDA instead of the LDA) is an evidence for quite strong dependence of this quantity on lattice parameters. As to location of the extrema, due to the relaxation the CBM has moved from \textbf{k}=(0.652,0.585,0.585) to \textbf{k}=(0.673,0.579,0.579). The VBM found in Ref.~\onlinecite{Wang_Cagin_PRB_2007} (at \textbf{k}=(0.650,0.584,0.584) before the relaxation and at \textbf{k}=(0.662,0.584,0.584) in the case of the relaxed structure) can be associated with the extremum C of the LDA calculations mentioned above. On the contrary, the LDA extremum X has ``appeared'' in  Ref.~\onlinecite{Wang_Cagin_PRB_2007} only in the relaxed bismuth telluride at \textbf{k}=(0.531,0.348,0.348) and as a second-highest VBM that is 26.7 meV lower than the VBM.

At fixed crystal-structure parameters, the band gap and the extrema positions vary also with the band-structure calculation method as it can be traced on the example of LDA-based calculations. In Ref.~\onlinecite{Kioupakis_PRB_2010}, with the use of the plane-wave \emph{ab initio} pseudopotential (PWP) method\cite{Ihm_JPhysC_1979} and Troullier-Martins\cite{Troullier_PRB_1991} pseudopotentials a value of 87 meV has been obtained for the same experimental structure parameters as those used in Ref.~\onlinecite{Youn_Freeman_PRB_2001} (61 meV). The VBM and the CBM have been found at \textbf{k}=(0.54, 0.37, 0.37) and at \textbf{k}=(0.68, 0.58, 0.58), respectively (see Fig.~\ref{fig2}). The second-highest VBM is located at \textbf{k}=(0.67, 0.58, 0.58) and 11 meV lower in energy than the VBM. The second-lowest CBM is along the $\Gamma-Z$ line and 62 meV higher than the CBM.

Recently, one more LDA study has been performed\cite{KKR_LDA_Yavorsky_PRB_2011} with the experimental crystal-structure data reported in Ref.~\onlinecite{Wyckoff}. The authors of Ref.~\onlinecite{KKR_LDA_Yavorsky_PRB_2011} used the screened Korringa-Kohn-Rostoker method in the atomic sphere approximation within the LDA of Ref.~\onlinecite{Vosko_CJP_1980}. They have found that the band gap is of 105 meV, and the VBM location is at \textbf{k}=(0.517, 0.366, 0.366) with $M=6$, while the CBM is on the $\Gamma-\mathrm{Z}$ line at \textbf{k}=(0.173, 0.173, 0.173) with $M=2$. Also, from the results presented in Ref.~\onlinecite{KKR_LDA_Yavorsky_PRB_2011}, one can gain insight into an effect of a compression of the lattice in the plane perpendicular to the hexagonal $c_H$ axis on the extrema locations and the band gap. The authors modeled the compression by replacing the lattice parameter $a_H$ of Bi$_2$Te$_3$ with that of Sb$_2$Te$_3$. Such a replacement causes a certain shift of the extrema without changing their multiplicity and an increase of the band gap up to 129 eV.

In Ref.~\onlinecite{Kioupakis_PRB_2010}, in addition to the LDA calculations the first treatment of $GW$ corrections to the LDA band structure of bismuth telluride has been done (see also Ref.~\onlinecite{Yazyev_PRB_R_2012}). The authors of Ref.~\onlinecite{Kioupakis_PRB_2010} have shown that the $GW$ corrections increase the band gap up to 0.17 eV. This means that, similar to the full structure optimization, these corrections have doubled the gap. In these LDA+$GW$ calculations, the VBM and the CBM are located away from the symmetry lines [at \textbf{k}=(0.66, 0.58, 0.58) and at \textbf{k}=(0.67, 0.58, 0.58), respectively] and have the multiplicity $M=6$ (see Fig.~\ref{fig2}). The second-highest VBM that is merely 1 meV lower than the VBM was found at \textbf{k}=(0.55, 0.38, 0.38). The second-lowest CBM that is along the $\Gamma-\mathrm{Z}$ line is located more then 75 meV higher in energy. As compared with Ref.~\onlinecite{Kim_Freeman_PRB_2005}, LDA+$GW$ values of the effective-mass parameters calculated in Ref.~\onlinecite{Kioupakis_PRB_2010} demonstrate closer agreement with the experiment in the case of the in-plane components found for the VBM.

The most of the mentioned theoretical results on the ``true'' positions of the extrema and the corresponding band-gap values are shown in Fig.~\ref{fig2}. As is clearly seen from the figure, the LDA-based calculations yield quite close but different locations of the corresponding band extrema and unanimously predict an indirect band gap (except for the case of the $GW$ calculations of Ref.~\onlinecite{Kioupakis_PRB_2010}, where a direct gap is also possible due to the presence of two nearly degenerate maxima of the valence band). In contrast, the GGA calculations of Ref.~\onlinecite{Wang_Cagin_PRB_2007} unambiguously point out that bismuth telluride possesses a direct band gap. However, its value is unexpectedly small. As to the multiplicity of the ``true'' extrema, only in Ref.~\onlinecite{KKR_LDA_Yavorsky_PRB_2011} (not shown in Fig.~\ref{fig2}) the CBM is on the $\Gamma-\mathrm{Z}$ line, which leads to $M=2$ that disagrees with experiment. Thus, summing up all the above theoretical results, one can infer that the positions of the extrema, the respective effective-mass parameters, and the band-gap value along with its character (direct or indirect) vary substantially with approximations to the XC functional, method for band structure calculations, and crystal-structure parameters.

Experimentally, the band gap in bismuth telluride has been determined by different methods. In Ref.~\onlinecite{Austin_PPS_1958}, optical measurements have led to an indirect band gap with the zero-temperature extrapolated value of 0.16 eV.  Resistivity measurements done in Ref.~\onlinecite{Li_JAP_1961} have revealed the gap of 0.17 eV. The authors of Ref.~\onlinecite{Sehr_JPChSol_1962} have found a thermal band gap of 0.15 eV and an optical band gap of 0.17 eV as obtained at 85$^{\circ}$K by Moss' criterion. In more recent experimental study,\cite{Thomas_PRB_1992} optical measurements at 10$^{\circ}$K have been performed, and a probably indirect band gap of $150\pm20$ meV and a probably direct gap of $220\pm20$ meV have been found.

Over the last several years, a huge number of experimental studies caused by the unique surface properties of bismuth telluride have been done to examine its surface electron structure by angle resolved photoemission spectroscopy (ARPES). From some of them, one can gain an information about the bulk band structure. For example, in Ref.~\onlinecite{Hsieh_PRL_2009} ARPES measurements have been done along several lines parallel to the Z$-$U direction and lying in the mirror plane. These lines have been chosen to contain positions of the extrema as obtained within FLAPW-GGA calculations performed with the optimized lattice parameters and atomic positions taken from Ref.~\onlinecite{Wang_Cagin_PRB_2007}. In these calculations, the VBM has been found at the same location as that obtained in Ref.~\onlinecite{Youn_Freeman_PRB_2001}. The authors of Ref.~\onlinecite{Hsieh_PRL_2009} came to the conclusion that the experimentally observed VBM location is in close agreement with calculations and, as a consequence, that Bi$_2$Te$_3$ has an indirect band gap with a low-limit estimate of $150\pm50$ meV.

In Ref.~\onlinecite{Chen_Science_2009}, within an ARPES study of $n$- and $p$-type doped bismuth telluride it was found that the band gap is of 0.165~eV, at that ARPES measurements of band dispersions along $\bar{\Gamma}-\bar{\mathrm{M}}$ show a minimum of the conduction band at the $\bar{\Gamma}$ point. This minimum is observed practically at the same energy as a bottom of a slightly blurred convex border, where the surface state ``touches'' with the conduction band, which forms an additional (not clearly seen) minimum (see also Ref.~\onlinecite{Henk_PRL_2012}). A precursor of such a touch is the bend of the surface-state dispersion and further ``opening up'' of the warped constant-energy contours of the surface state on the $\bar{\Gamma}-\bar{\mathrm{M}}$ line (see also Ref.~\onlinecite{Alpichshev_PRL_2010}). As to the VBM, in Ref.~\onlinecite{Chen_Science_2009} this maximum seems to be much closer to $\bar{\Gamma}$ on $\bar{\Gamma}-\bar{\mathrm{M}}$ than it comes from Ref.~\onlinecite{Hsieh_PRL_2009}. A similar situation is observed in Ref.~\onlinecite{Ogawa_PRB_2012}, where the surface electronic structure of an $n$-type doped Bi$_2$Te$_3$ has been examined by the ARPES. The photoemission intensity plot along $\bar{\Gamma}-\bar{\mathrm{M}}$ and derived dispersion curves reported in this work show that the CBM is located away from $\bar{\Gamma}$ at the parallel momentum $k_{\bar{\Gamma}\bar{M}}\sim0.11$ \AA$^{-1}$ that is slightly larger than that of the mentioned bend of the surface state. As to the VBM, it appears to be not far away from the momentum of the CBM.

The cited experimental results give a quite small scatter of the band-gap values and are in favor of the indirect character of the band gap. As to the band dispersion in the vicinity of the extrema and their positions in the mirror plane of the BZ, the effective-mass parameters and the multiplicity of the extrema are known form Shubnikov-de Haas investigations done in Refs.~\onlinecite{Koehler_PSS_1976_1} and \onlinecite{Koehler_PSS_1976_2} with $n$- and $p$-doped samples (correspond to +30.5 meV and -23.8 meV, respectively). A purposeful study of the bulk band-gap edges as, e.g., in Ref.~\onlinecite{Nechaev_PRBR_2013}, where an ARPES study of bismuth-selenide band structure has been performed by probing a large fraction of \textbf{k} space on a dense grid of emission angles and photon energies, has not been done so far.

In this paper, we report a theoretical study of the dispersion of the highest valence and the lowest conduction bands in large fraction of the BZ of bulk bismuth telluride. On the same footing, we consider all the aforementioned factors which can lead to changes in the extrema locations and the band-gap value. We show how the atomic positions, the approximation to the DFT-XC functional, the $GW$ many-body corrections to the DFT states affect the extrema location in \textbf{k} space, the effective-mass parameters calculated for the VBM and the CBM, and the band gap.

\section{\label{sec:calc_details}Calculation details}

Similar to Ref.~\onlinecite{Nechaev_PRBR_2013}, in our \textit{ab initio} calculations we employ the FLAPW method as implemented in the FLEUR code\cite{FLEUR} within both the LDA of Ref.~\onlinecite{LDA_CA_PZ} and the GGA of Ref.~\onlinecite{GGA_PBE} for the XC functional. The ground-state calculations were carried out with the use of a plane-wave cutoff of $k_{max}=4.5$ bohr$^{-1}$, an angular momentum cutoff of $l_{max}=10$, equal muffin-tin radii of 2.79 \AA\, for Bi and Te, and a $7\times7\times7$ $\Gamma$-centered $\mathbf{k}$-point sampling of the BZ. The FLAPW basis has been extended by conventional local orbitals\cite{Singh_PRB_1991, Sjoestedt_SSC_2000} to treat semi-core $d$-states ($4d$ for Te and $5d$ for Bi). The energy cutoff between core and valence states was put at -1.4 Ha, what corresponds to 78 valence electron in the considered energy window in a rhombohedral Bi$_2$Te$_3$. To more accurately describe high-lying unoccupied states,\cite{Krasovskii_PRB_1997} one local orbital per angular momentum up to $l = 3$ was included for each atom. In all calculations, the Fermi level was placed in the middle of the band gap.

Many-body corrections to GGA- and LDA-states are found within the one-shot $GW$ approach as realized by the SPEX code.\cite{GW_SPEX} The spin-orbit interaction was included into the $GW$ calculations already at the level of the reference one-particle band structure.\cite{Sakuma_PRB_2011} The dielectric matrix was evaluated within the random-phase approximation and represented with the use of the mixed product basis,\cite{GW_SPEX,Kotani_SSC_2002} where we chose an angular momentum cutoff in the muffin-tin spheres of 4 and a linear momentum cutoff of 3.5 bohr$^{-1}$. The $GW$ calculations were performed with the number of unoccupied bands $N_b=252$ and less dense Monkhorst-Pack grid ($4\times4\times4$) than in the case of the DFT calculations. The detailed study of the convergence with respect to the number of unoccupied states\cite{Friedrich_PRBR_2011} has revealed that the band gap at the $\Gamma$ point decreases with increasing $N_b$, and the parameters indicated above ensure the $\Gamma$-point band gap converged within 27 meV. A move to a denser \textbf{k}-pint grid at the fixed $N_b=252$ causes the decrease of the $\Gamma$-point gap by 21 meV.

We investigate the behavior of the valence and conduction bands in the mirror plane (or, more precisely, in the part of this plane shown in Fig.~\ref{fig1}(b) by the dotted-line rectangle), which is sampled by a dense equidistant mesh composed of 225 {\bf k} points (900 {\bf k} points in the case of the DFT calculations). For each point a separate $GW$ calculation was performed. On the basis of this mesh, we made a guess of the extrema locations which ware successively defined more accurately by performing $GW$ calculations on a finer mesh in the vicinity of the guess.

We consider three sets of atomic positions for Bi and Te atoms in the rhombohedral crystal structure [see Fig.~\ref{fig1}(a)] with experimental lattice parameters ($a_H=4.3853$ \AA\, and $c_H=30.487$ \AA) taken from Ref.~ \onlinecite{Wyckoff}. The fist one is labeled as `0' and corresponds to atomic positions reported in Ref.~\onlinecite{Wyckoff} (Te$^\mathrm{I}$ at (0.000, 0.000, 0.000), Te$^\mathrm{II}$ at ($\pm\mu$, $\pm\mu$, $\pm\mu$) with $\mu=0.212$, and Bi at ($\pm\nu$, $\pm\nu$, $\pm\nu$) with $\nu=0.400$ as presented in real-lattice-vector coordinates\cite{remark_coords}). The second and third sets which we label as `I' and `II' were obtained during a relaxation procedure optimizing the atomic positions at fixed volume until forces became less than $1.0\times10^{-3}$ Ha/bohr within the LDA and GGA calculations, respectively. After such a relaxation procedure, within the LDA calculations we have $\mu=0.2101$ and $\nu=0.3994$. In the case of the GGA calculations, we arrived at $\mu=0.2089$ and $\nu=0.4000$. It is worth noting that all these relaxed positions fall in the error range of the experiment presented in Ref.~\onlinecite{Nakajima_JPCS_1963} ($\mu=0.2097\pm0.0009$ and $\nu=0.4000\pm0.0007$). Moreover, the lattice parameters of Ref.~\onlinecite{Wyckoff} are in the error range of the experimental values $a_H=4.386\pm0.005$ \AA\, and $c_H=30.497\pm0.020$ \AA\, reported in Ref.~\onlinecite{Nakajima_JPCS_1963}. This means that the considered three sets of atomic positions cover the available experimental data on crystal structure of Bi$_2$Te$_3$.

To estimate the effective-mass tensor parameters ($\alpha_{ij}$), the valence- and conduction-band energy near the extremum points, which are lying in the $yz$ mirror plane in $\mathbf{k}$ space, is approximated by the expression
$$%\begin{equation}\label{approximation}
E(\mathbf{k})=E_0+\mathbf{v}\cdot \mathbf{k} + \mathbf{k}\cdot \mathbf{Q} \cdot \mathbf{k}
$$%\end{equation}
with $Q_{ij}=\hbar^2\alpha_{ij}/2m_e$. The matrix elements $Q_{xy}$ and $Q_{xz}$ are put at zero. The rest 8 parameters are found within the least squares method by fitting the band energy $E(\mathbf{k})$ calculated on an additional 29 $\mathbf{k}$-point mesh centered at the extremum point ($\Delta k_i=\pm0.0025$ a.u.$^{-1}$). The principle angle of the energy ellipsoid in the mirror plane with respect to the $y$-axis is defined as
$$
\theta_{yz}=\frac{1}{2}\arctan\left(\frac{2\alpha_{yz}}{\alpha_{zz}-\alpha_{yy}}\right).
$$

\section{\label{sec:results}Results}

\subsection{\label{subsec:GGA_based}GGA-based calculations}

\begin{figure*}[tbp]
\centering
 \includegraphics[angle=0,scale=1.9]{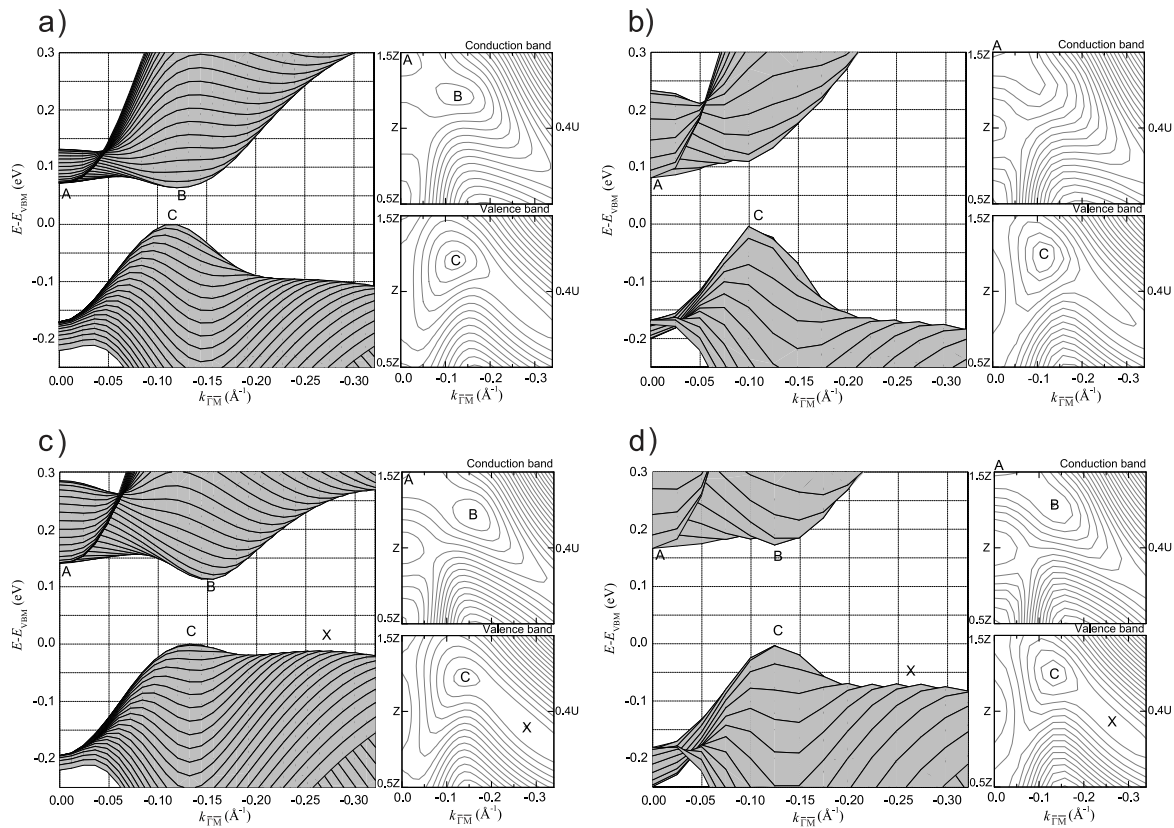}
\caption{Each panel contains projections on the $\bar{\Gamma}-\bar{M}$ direction of the two-dimensional BZ and contour plots of the lowest conduction band and the uppermost valence band in the mirror plane. The presented results are obtained for the unrelaxed (the set `0', upper row) and relaxed (the set `II', lower row) atomic positions without (left column) and with (right column) the $GW$ corrections to the GGA band structure. The letters `A', `B', `C', and `X' mark extrema discussed in the text.}\label{fig3}
\end{figure*}

We start with the GGA band structure found with the experimental lattice parameters\cite{Wyckoff} and two sets of atomic positions (`0' and `II'). Figs.~\ref{fig3}(a) and \ref{fig3}(c) show the projections of the lowest conduction-band and the uppermost valence-band dispersions in the mirror plane on the $\bar{\Gamma}-\bar{M}$ direction of the two-dimensional BZ. The figures also contain the contour plots of the mentioned bands. First, we note that the band structures presented in Figs.~\ref{fig3}(a) and \ref{fig3}(c) reflect an influence of the small displacement of Te$^{\mathrm{II}}$. Upon this displacement that changes the distance between quintuple layers, the ``fundamental'' band gap becomes larger. Both in the unrelaxed case [the set `0', \ref{fig3}(a)] and in the relaxed case [the set `II', \ref{fig3}(c)], the role of the CBM is played by the extremum B, while the VBM is presented by the extremum C (both have the multiplicity $M=6$). The energy difference between these extrema changes from 65 meV to 114 meV upon relaxing the atomic positions (see also Table~\ref{tab:table5}). The second-lowest CBM marked as A is located 8 meV higher in energy in the unrelaxed Bi$_2$Te$_3$ and 28 meV in the relaxed one. The location of the extrema in the mirror plane is clearly presented in the contour plots which are also shown in Figs.~\ref{fig3}(a) and \ref{fig3}(c). For the unrelaxed atomic positions, the local conduction-band minimum A is at \textbf{k}=(0.264, 0.264, 0.264) [on the $\Gamma-Z$ line], the CBM is located at \textbf{k}=(0.656, 0.583, 0.583), and the VBM is found at \textbf{k}=(0.650, 0.580, 0.580). In the relaxed case, these extrema are located at \textbf{k}=(0.248, 0.248, 0.248), \textbf{k}=(0.669, 0.578, 0.578), and \textbf{k}=(0.667, 0.586, 0.586), respectively, which indicate a certain shift of the extrema.

The extremum X appears at \textbf{k}=(0.556, 0.397, 0.397) in the relaxed case only and is 12 meV lower than the extremum C, which reflects an opposite relation of these extrema as compared with the LDA calculations mentioned in Sec.~\ref{sec:introduction} (see also Fig.~\ref{fig2}). This situation is close to that reported in Ref.~\onlinecite{Wang_Cagin_PRB_2007}, where GGA calculations have been performed with the FLEUR code too. However, in Ref.~\onlinecite{Wang_Cagin_PRB_2007} the band gap is substantially smaller than that in our study. As our analysis has shown, such a big difference may be caused by the treatment of the quite shallow semi-core $d$ states of Bi in Ref.~\onlinecite{Wang_Cagin_PRB_2007} as valence states without resorting to the local orbitals.

%=============================================================================================================
%=============================================================================================================

%=============================================================================================================
\begin{table}
\caption{\label{tab:table1} The effective-mass tensor parameters for the VB extrema as compared with the
experimental ones taken from Ref.~\onlinecite{Koehler_PSS_1976_1}. The results of the GGA-based calculations are presented for different sets (indicated in the parentheses) of the atomic positions.}
\begin{ruledtabular}
\begin{tabular}{lcccccc}
                  Calculations   &  Extremum   & $\alpha_{xx}$ & $\alpha_{yy}$ & $\alpha_{zz}$ & $\alpha_{yz}$ & $\theta_{yz}$ \\
  \hline
  GGA(0)        &   C    & 51.1        &  9.0         & 14.5         & -2.0          &   -18$^{\circ}$       \\
  GGA(0)+$GW$   &   C    & 44.4        &  14.2        & 13.3         & -1.8          &    38$^{\circ}$       \\
  GGA(II)       &   C    & 53.4        &  4.0         & 10.4         &  0.2          &     1$^{\circ}$      \\
  GGA(II)+$GW$  &   C    & 54.2        &  10.3        & 11.1         &  1.1          &    36$^{\circ}$       \\
  GGA(II)       &   X    & 47.9        &  4.8         & 6.4          &  5.1          &    41$^{\circ}$       \\
  GGA(II)+$GW$  &   X    & 26.3        &  4.1         & 5.4          &  4.4          &    41$^{\circ}$       \\
  \multicolumn{2}{l}{Experiment ($\pm10\%$)}   & 32.5        &  4.81        & 9.02         &  4.15          &    31.5$^{\circ}$       \\

\end{tabular}
\end{ruledtabular}
\end{table}
%=============================================================================================================

%=============================================================================================================
\begin{table}
\caption{\label{tab:table2} The effective-mass tensor parameters for the CB extrema as compared with the
experimental ones taken from Ref.~\onlinecite{Koehler_PSS_1976_2}. The results of the GGA-based calculations are presented for different sets (indicated in the parentheses) of the atomic positions.}
\begin{ruledtabular}
\begin{tabular}{lcccccc}
                  Calculations   &  Extremum   & $\alpha_{xx}$ & $\alpha_{yy}$ & $\alpha_{zz}$ & $\alpha_{yz}$ & $\theta_{yz}$ \\
  \hline
  GGA(0)         &   A    &  3.2        &  3.2         &  1.5        &   0.0       &    0$^{\circ}$      \\
  GGA(0)+$GW$    &   A    &  5.4        &  5.3         &  1.7        &   0.0       &    0$^{\circ}$      \\
  GGA(II)        &   A    &  3.5        &  3.5         &  3.1        &   0.0       &    0$^{\circ}$      \\
  GGA(II)+$GW$   &   A    &  5.3        &  5.3         &  3.4        &   0.0       &    0$^{\circ}$      \\
  GGA(0)         &   B    &  60.2       &  4.1         &  14.9       &   0.6       &    3$^{\circ}$        \\
  GGA(II)        &   B    &  71.4       &  7.7         &  11.3       &   2.9       &    29$^{\circ}$       \\
  GGA(II)+$GW$   &   B    &  62.9       &  8.0         &  10.7       &   3.7       &    35$^{\circ}$       \\
  \multicolumn{2}{l}{Experiment ($\pm10\%$)}   &  46.9       &  5.92        &  9.50       &   4.22      &    33.5$^{\circ}$       \\

\end{tabular}
\end{ruledtabular}
\end{table}
%=============================================================================================================

\begin{figure}[tbp]
\centering
 \includegraphics[angle=0,scale=0.8]{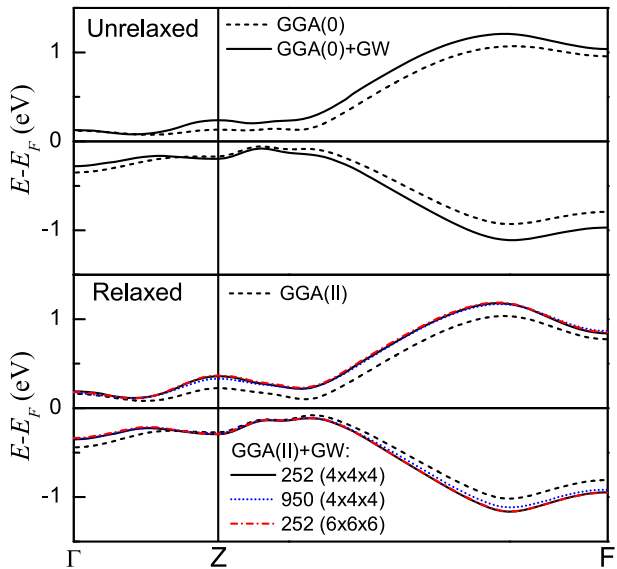}
\caption{(Color online) Valence and conduction bands as found within the GGA and the GGA+$GW$ for the unrelaxed (top) and relaxed
(bottom) cases. For the relaxed case, the GGA+$GW$ results obtained with different numbers of unoccupied bands and \textbf{k}-point grids (indicated in parentheses) are shown.}\label{fig4}
\end{figure}

Fig.~\ref{fig4} shows the GGA band structure along the $\Gamma-$Z$-$F line. As seen in the figure, upon the displacement of the Te$^{\mathrm{II}}$ atom, the band gap along these lines increases from 132 meV to 161 meV. The relaxation of the atomic positions enlarges notably the band gap at the $\Gamma$-point too. It reflects an increase of the ``penetration'' of the conduction and the valence bands into each other near the $\Gamma$ point and, as a consequence, a broadening of the band-inversion region (see, also, Ref.~\onlinecite{Yazyev_PRB_R_2012}). On the whole, as compared with the experimental data our GGA calculations do not demonstrate the experimentally observed character and value of the bulk band gap, though the correct multiplicity of the extrema is reproduced. As to the effective-mass tensor parameters found for the VBM (see Table~\ref{tab:table1}) and the CBM (see Table~\ref{tab:table2}), similar to Ref.~\onlinecite{Wang_Cagin_PRB_2007} the obtained GGA values are quite far from their experimental counterparts, except for those for the extremum X and the extremum B (the CBM) which appear in the GGA calculations performed for the relaxed atomic positions.

In Figs.~\ref{fig3}(b) and \ref{fig3}(d), we show the highest valence and the lowest conduction bands as obtained with the $GW$ corrections to the GGA bands for the considered sets of atomic positions. Regarding these figures, it is worth noting that changes caused by taking into account many-body corrections occur in both \textbf{k}-space locations of the extrema and their relative positions on the energy scale. In the unrelaxed case [Fig.~\ref{fig3}(b)], the extremum B disappears, the extremum A moves slightly towards the $\Gamma$-point [\textbf{k}=(0.206, 0.206, 0.206)] and now plays the role of the CBM with the multiplicity $M=2$, and, finally, the extremum C [\textbf{k}=(0.655, 0.601, 0.601)] remaining in the capacity of the VBM becomes more pronounced. In the relaxed case, after inclusion of the $GW$ corrections, we have found A at \textbf{k}=(0.209, 0.209, 0.209), B at \textbf{k}=(0.685, 0.606, 0.606), C at \textbf{k}=(0.677, 0.600, 0.600), and X at \textbf{k}=(0.556, 0.402, 0.402). Here, the CBM and the VBM are presented by A and C, respectively. As compared with the GGA dispersions, the VBM gets more prominent with respect to X that, in turn, becomes less evident [see Fig.~\ref{fig3}(d)].The extremum B is not so deep as in Fig.~\ref{fig3}(c) and on the energy scale is of 12 meV higher than the CBM (the extremum A) only.

The ``fundamental'' band gap, which can be clearly seen in Figs.~\ref{fig3}(b) and \ref{fig3}(d) as the energy interval between A and C, has the value of 76 meV in the unrelaxed case and amounts to 156 meV in the relaxed case, where the second-highest VBM (the extremum X) is 75 meV lower in energy than C (see Table~\ref{tab:table5}). As seen in Fig.~\ref{fig4}, similar to the conventional semiconductor systems (see, e.g., Ref.~\onlinecite{GW_semiconds}) in Bi$_2$Te$_3$ the $GW$ corrections enlarge the bulk band gap mainly by ``moving'' the conduction band away from the valence band on the energy scale, except the vicinity of the $\Gamma$-point. This exception is caused by the band inversion near the center of the BZ. In contrast to Bi$_2$Se$_3$, where many-body corrections leads to a shift of the VBM from the location in the mirror plane to $\Gamma$,\cite{Yazyev_PRB_R_2012, Nechaev_PRBR_2013} in bismuth telluride the band inversion does not induce such a crucial rearrangement of VB extrema upon moving the considered bands apart. However, to some extent it is applied to the conduction band. Actually, in the $GW$ calculations the CBM is presented by the extremum A that lies on the $\Gamma$-Z line, while the extremum B (the GGA CBM) disappears (in the unrelaxed case) or notably increases (in the relaxed case) its energy as compared with the extremum C. What unites the two mentioned topological insulators is that due to the band inversion the band gap at the $\Gamma$ point becomes smalle (see Fig.~\ref{fig4}).

%=============================================================================================================
%=============================================================================================================

It is worth noting that in the relaxed case, due to a slight shift of the locations of B and C in the mirror plane, these two extrema form a direct band gap that is larger than an indirect gap between A and C. Along the $\Gamma$-Z-F line, the $GW$ band gap has the value of 160 meV for the unrelaxed case and of 227 meV in the relaxed one. This means that as compared with the case of the experimental crystal structure taken from Ref.~\onlinecite{Wyckoff} many-body effects have more profound influence on the band gap in the case of the atomic positions of Ref.~\onlinecite{Nakajima_JPCS_1963}. It is significant that, in contrast to the $\Gamma$-point band gap (see Sec.~\ref{sec:calc_details}), the $GW$ band gap along the $\Gamma$-Z-F line decreases slightly with the increase in the number of unoccupied bands $N_b$. In the relaxed case, its value fell to 225 meV at $N_b=950$ with the unchanged \textbf{k}-point grid (see Fig.~\ref{fig4}). At fixed $N_b=252$, a denser \textbf{k}-point grid ($6\times6\times6$) causes the decrease to 220 meV only.

The GGA-based $GW$ calculations predicting the CBM to be located on the $\Gamma$-Z line are not in agreement with the available experimental data on the multiplicity and the effective-mass tensor parameters (see Table~\ref{tab:table2}). In this respect, the extremum B surviving in the relaxed case is more fit for the role of the CBM. Since the energy distance between A and B is within the interval of convergence and notably smaller than the experimental Fermi energy of 30.5 meV (measured from the bottom of the CB) caused by relatively high $n$-type doping done in Ref.~\onlinecite{Koehler_PSS_1976_2}, argumentations reported in Refs.~\onlinecite{Youn_Freeman_PRB_2001} and \onlinecite{Kim_Freeman_PRB_2005} may hold true. The point is that such a doping may result in connection of these extrema in a combined Fermi surface with $\theta_{yz}$ which is quite well reproduced in the GGA(II)+$GW$ calculations for the extremum B. Nevertheless, at any electron doping the GGA(II)+$GW$ results disconfirm the six-valley model\cite{Drabble_PPSL_1958} for the conduction band. However, the obvious atomic-position dependence of the relative positions of A and B on the energy scale indicates that due to, e.g., temperature effect the CBM can be already presented by B.

As regards to the valence band and its extrema, the effective-mass tensor parameters (the in-plane and out-of-plane components) found for X are in good agreement with experiment, while the angle $\theta_{yz}$ is reproduced better in the case of the extremum C. The latter is the calculated VBM that corroborates the six-valley model of Ref.~\onlinecite{Drabble_PPSL_1958} for the valence band. The extremum X is an extensive local maximum which can be connected with C in a combined Fermi surface in the case of quite high hole doping only.

\subsection{\label{subsec:LDA_based}LDA-based calculations}

%=============================================================================================================
\begin{figure*}[tbp]
\centering
 \includegraphics[angle=0,scale=1.85]{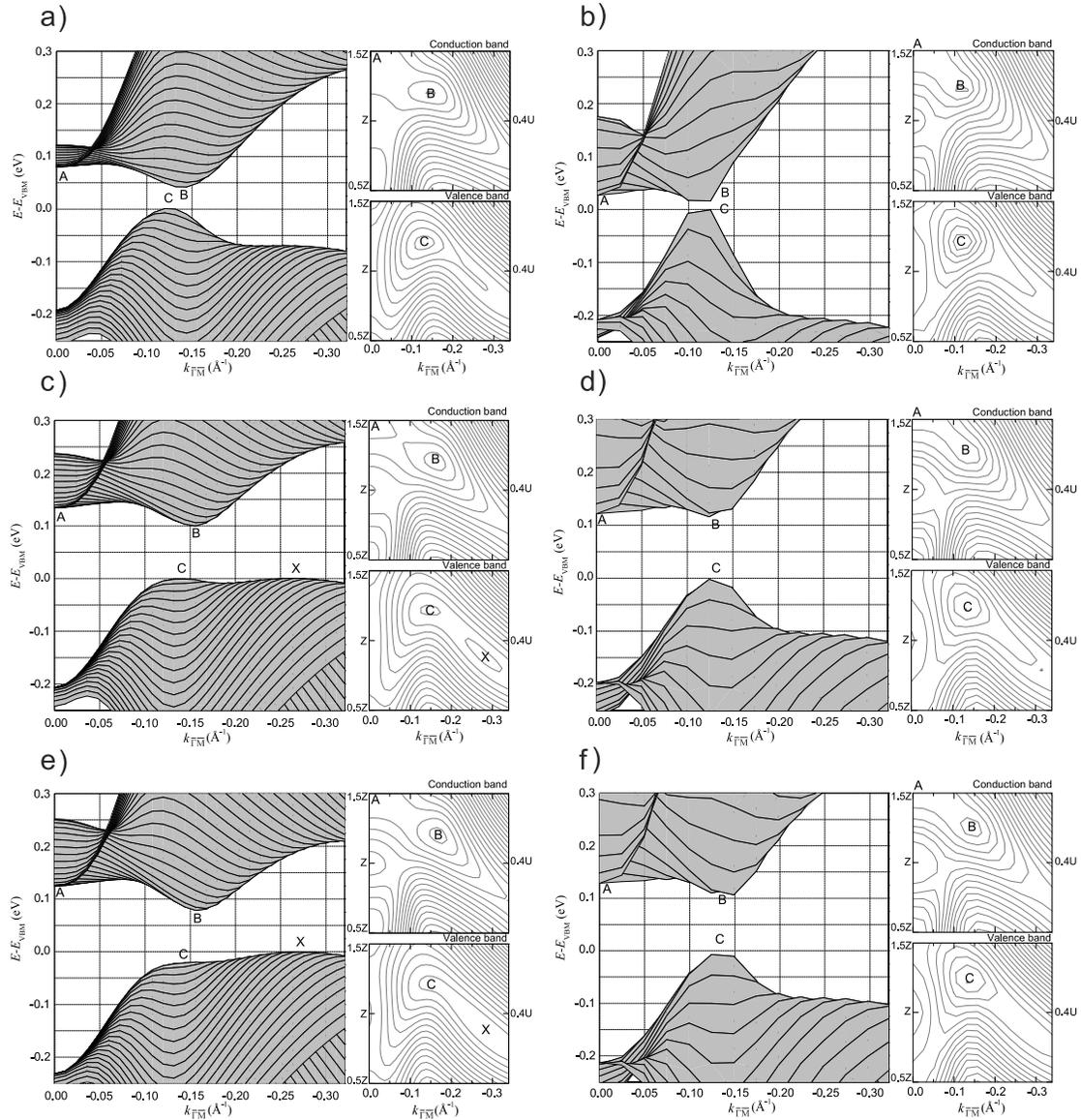}
\caption{Same as in Fig.~\ref{fig3}, but for the LDA-based calculations. Upper, middle, and lower rows correspond to the sets of atomic positions labeled in the text as ``0'', ``I'', and ``II'', respectively.}\label{fig5}
\end{figure*}
%=============================================================================================================
%=============================================================================================================
\begin{table}
\caption{\label{tab:table3} Same as in Table \ref{tab:table1}, but in the case of the LDA-based calculations and as compared with other \textit{ab initio} calculations.}
\begin{ruledtabular}
\begin{tabular}{lcccccc}
      Calculations   &  Extremum   & $\alpha_{xx}$ & $\alpha_{yy}$ & $\alpha_{zz}$ & $\alpha_{yz}$  & $\theta_{yz}$          \\
  \hline
  LDA(I)                        &   C    &    55.7       &  2.6         & 10.8          &  1.2           &    8$^{\circ}$        \\
  LDA(I)+$GW$                   &   C    &    77.3       &  12.2        & 16.1          &  2.1           &    24$^{\circ}$        \\
  LDA(II)+$GW$                  &   C    &    85.3       &  10.1        & 17.2          &  1.9           &    14$^{\circ}$        \\
  LDA(I)                        &   X    &    50.1       &  5.0         & 6.7           &  5.4           &    41$^{\circ}$        \\
  LDA(II)                       &   X    &    63.1       &  5.4         & 7.5           &  5.9           &    41$^{\circ}$        \\
\hline
  \multicolumn{2}{l}{Experiment ($\pm10\%$)}   &    32.5       &  4.81        & 9.02          &  4.15          &    31.5$^{\circ}$        \\
\hline
  PWP LDA+$GW$\footnotemark[2]  &   C    &    47.33      &  9.94        & 14.61         &  -1.25         &    -14.0$^{\circ}$        \\
  FLAPW sX-LDA\footnotemark[1]  &   X    &    39.5       &  3.8         & 5.2           &  6.2           &    41$^{\circ}$        \\
  PWP LDA\footnotemark[2]       &   X    &    56.93      &  4.84        & 6.64          &  5.21          &    40.1$^{\circ}$        \\
  PWP LDA+$GW$\footnotemark[2]  &   X    &    45.87      &  7.46        & 10.17         &  5.16          &    37.6$^{\circ}$        \\
\end{tabular}
\footnotetext[1]{From Ref.~\onlinecite{Kim_Freeman_PRB_2005}}
\footnotetext[2]{From Ref.~\onlinecite{Kioupakis_PRB_2010}}
\end{ruledtabular}
\end{table}
%=============================================================================================================

%=============================================================================================================
\begin{table}
\caption{\label{tab:table4} Same as in Table \ref{tab:table2}, but in the case of the LDA-based calculations and as compared with other \textit{ab initio} calculations.}
\begin{ruledtabular}
\begin{tabular}{lcccccc}
            Calculations   &  Extremum   & $\alpha_{xx}$ & $\alpha_{yy}$ & $\alpha_{zz}$ & $\alpha_{yz}$ & $\theta_{yz}$ \\
  \hline
  LDA(I)                        &   B    &    73.4       &  7.7         & 11.8          &  3.3           &    29$^{\circ}$        \\
  LDA(I)+$GW$                   &   B    &    80.3       &  9.7         & 13.8          &  4.3           &    32$^{\circ}$        \\
  LDA(II)                       &   B    &    88.4       &  8.9         & 11.3          &  4.2           &    37$^{\circ}$        \\
  LDA(II)+$GW$                  &   B    &    98.9       &  11.9        & 16.9          &  4.5           &    30$^{\circ}$        \\
\hline
  \multicolumn{2}{l}{Experiment ($\pm10\%$)}   &   46.9       &  5.92        & 9.50          &  4.22          &    33.5$^{\circ}$        \\
\hline
  FLAPW sX-LDA\footnotemark[1]  &   B    &    52.2       &  8.0         & 7.3           &  3.8           &    -42.4$^{\circ}$        \\
  PWP LDA\footnotemark[2]       &   B    &    82.25      &  7.96        & 10.39         &  3.72          &    36.0$^{\circ}$        \\
  PWP LDA+$GW$\footnotemark[2]  &   B    &    57.18      &  8.93        & 12.50         &  1.74          &    22.1$^{\circ}$        \\

\end{tabular}
\footnotetext[1]{From Ref.~\onlinecite{Kim_Freeman_PRB_2005}}
\footnotetext[2]{From Ref.~\onlinecite{Kioupakis_PRB_2010}}
\end{ruledtabular}
\end{table}
%=============================================================================================================

Now we turn to our LDA-based calculations and consider first LDA band structures as obtained with different sets of atomic positions. Fig.~\ref{fig5}(a) shows the projections and contour plots of the lowest conduction band and the uppermost valence band in the mirror plane, which were found for the experimental atomic positions. A comparison with the respective GGA results presented in Fig.~\ref{fig3}(a) reveals how the change of approximation to the exchange-correlation functional affects the band gap and the band dispersions. The gap formed by the extrema B (the CBM) and C (the VBM) becomes notably smaller (see Table~\ref{tab:table5}), at that the energy intervals between the extrema A and B increases. It is worth noting that the locations of the VBM and the CBM in the mirror plane remain practically the same: B and C ware found at \textbf{k}=(0.656, 0.573, 0.573) and \textbf{k}=(0.646, 0.571, 0.571), respectively.

Similar to the GGA calculations presented in the previous subsection, upon relaxing atomic positions (see Fig.~\ref{fig5}(c), where the set `I' is used), the LDA band gap increases from 41 meV to 99 meV (see Table~\ref{tab:table5}). The extrema B and C, which as in the unrelaxed case are the CBM and the VBM, respectively, are slightly moved in \textbf{k}-space [B at \textbf{k}=(0.666, 0.574, 0.574), C at \textbf{k}=(0.665, 0.582, 0.582)]. As well as in the GGA case, the extremum X appears in the relaxed case only [at \textbf{k}=(0.551, 0.391, 0.391)]. However, in this case X has the same energy as C and, as a consequence, also can be considered as the VBM (note that in the GGA relaxed case X is merely 12 meV lower than C).

To make a direct comparison between GGA and LDA calculations for the relaxed atomic positions, we have calculated the LDA band structure for the set `II'. Figs. \ref{fig3}(c) and \ref{fig5}(e) clearly demonstrate that the conduction band undergos changes which are similar to those for the set `0': the extremum B becomes deeper and closer to the valence band, which reduces the band gap. However, as distinct from the set `0', but in agreement with other DFT-LDA calculations the fundamental band gap is formed by the extrema B and X (see Fig.~\ref{fig2}). (Here, we do not examine the extremum A, since in the LDA-based calculations this local minimum does not play a role of the CBM.) As to the extrema locations, we have found B at \textbf{k}=(0.665, 0.568, 0.568), C at \textbf{k}=(0.664, 0.569, 0.569), and X at \textbf{k}=(0.544, 0.381, 0.381).

By comparing the LDA and GGA calculations performed for different atomic positions, we clearly show the sensitivity of the band gap and the dispersion of the bands under study in the mirror plane. On the same footing, we demonstrate that in both approximations to the XC functional the small displacement of Te$^{\mathrm{II}}$ causes the substantial modifications of profile of the band-gap edges. The change of the approximation at fixed atomic positions also provokes differences in the band gap (see Table~\ref{tab:table5}) and the band dispersion (see the respective panels in Figs. \ref{fig3} and \ref{fig5}). As to the latter, from values of the effective-mass tensor parameters listed in Tables \ref{tab:table1}--\ref{tab:table4} it can be seen as well.

%=============================================================================================================
\begin{figure}[tbp]
\centering
 \includegraphics[angle=0,scale=0.53]{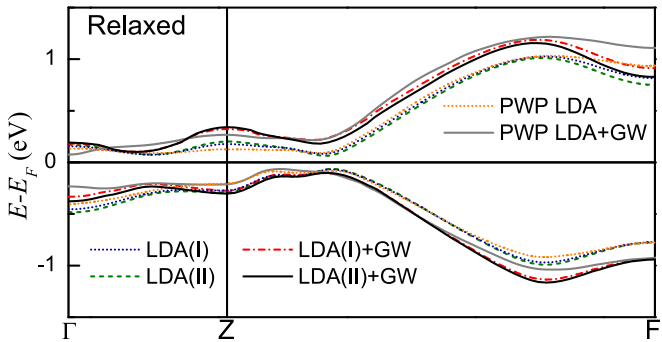}
\caption{(Color online) Valence and conduction bands as found within the LDA and the LDA+$GW$ for two relaxed sets of atomic positions. Curves marked by ``PWP'' reflect data taken from Ref.~\onlinecite{Yazyev_PRB_R_2012}. }\label{fig6}
\end{figure}
%=============================================================================================================

As in the GGA calculations, the effective-mass tensor parameters found within the LDA for the extremum X are closer to the experimental values than those for the extremum C. In the case of the set `II', the in-plane components of the effective-mass tensor parameters obtained within the LDA for the extremum B and, as a consequence, the angle $\theta_{yz}$ are in better agreement than in the GGA relaxed case. The out-of-plane component is farther from experiment than that obtained in the GGA. It is worth noting that the LDA effective masses calculated for B and X nicely match the LDA results of Ref.~ \onlinecite{Kioupakis_PRB_2010}, where the plane-wave pseudopotential method was used.

In Fig.~\ref{fig6}, we present the LDA band structure along the $\Gamma-$Z$-$F line as compared with that taken from Ref.~\onlinecite{Kioupakis_PRB_2010}. Note that along $\Gamma-$Z$-$F differences between the band dispersions in the LDA relaxed cases [LDA(I) and LDA(II)] are not so marked as in the mirror plane. Nevertheless, along this line the LDA(I) band gap is of 142 meV, while the LDA(II) one is of 129 meV. As compared to the LDA results of Ref.~ \onlinecite{Kioupakis_PRB_2010}, our results mainly differ by a bigger band gap at the BZ center. On the whole, by considering our LDA calculations in the light of the aforementioned experimental data, we can mark out the LDA(II) results which are notable for extrema hierarchy and their multiplicity in agreement with the experimental observations (though with smaller band-gap values than the experimental ones).

Our $GW$ calculations performed with the LDA reference one-particle band structure result in a strong reduction of the band gap in the case of the set `0'. (A solution of the quasiparticle equation taking into account off-diagonal matrix elements of the self-energy might improve such a behavior of the band-gap edges in this case,\cite{Aguilera_PRB_2013} since this small energy separation of the bands in the \textbf{k}-space region where hybridization is strong appears to be very sensitive to many-body renormalization.) This situation is too far from the experimental observations, and further we do not examine the extrema properties for this experimental\cite{Wyckoff} set of the atomic positions. Contrary, the both relaxed cases are characterized by an enlarged band gap, which is formed by the extrema B and C [see Figs. \ref{fig5}(d) and \ref{fig5}(f)]. The LDA(I)+$GW$ spectrum has C at \textbf{k}=(0.672, 0.594, 0.594) and B at \textbf{k}=(0.680, 0.602, 0.602). In the LDA(II)+$GW$ case, the extrema B and C are located at \textbf{k}=(0.677, 0.593, 0.593) and \textbf{k}=(0.671, 0.591, 0.591), respectively. It is worth noting here that, in spite of the fact that the extremum X is the VBM at the LDA level, in the $GW$ calculations the VBM is presented nevertheless by the extremum C as in the GGA(II)+$GW$ case, where on energy scale X lies notable lower than C. We show thus that, independently on the atomic positions (varying within the error of range of the available experimental data) and the approximation chosen for the XC functional (the LDA or the GGA), at the $GW$ level the VBM is presented by the extremum C. The effective-mass tensor parameters calculated for this extremum with taking into account $GW$ corrections are listed in Table~\ref{tab:table3}. As in Ref.~\onlinecite{Kioupakis_PRB_2010}, these parameters are quite far from the experiment.

As regards the CBM, in the LDA-based calculations the CBM is presented by the extremum B in all the cases. However, it is fair to say that in the LDA(I)+$GW$ case the extremum A is merely 3 meV higher than B. This means that, as well as in the GGA(II)+$GW$ case, with the doping used in the experiment\cite{Koehler_PSS_1976_2} (+30.5 meV) these two esxtrema are connected in a combined Fermi surface with $\theta_{yz}$ which for B is in good agreement with the experiment. Note that it also holds for the LDA(II)+$GW$ results which are characterised by A that is 21 meV higher than B. Keeping in mind that our LDA results are in consistent with those taken from Ref.~\onlinecite{Kioupakis_PRB_2010}, we would like to emphasize that at the $GW$ level there is an essential difference (see Tables \ref{tab:table3} and \ref{tab:table4}). It is also clearly seen in Fig.~\ref{fig6}, where the vicinity of the $\Gamma$ point is more demonstrative. In this figure, the band dispersions along $\Gamma$-Z-F line are shown. Along this line the band gap is of 207 meV in the LDA(I)+$GW$ case and of 203 meV in the LDA(II)+$GW$ case.\cite{band_gap_remark} Thus we can see that different realizations of the $GW$ corrections leads to substantially different behaviour of the band-gap edges on $GW$ level even if on the LDA level the reference band structures are quite similar. As a consequence, it will have an impact on characteristics of surface states forming the Dirac cone in the bulk band gap. (A detailed study of the effect of different approaches to the spin-orbit interaction in constructing many-body corrections on the quasiparticle spectrum can be found in Ref.~\onlinecite{Aguilera_PrComm_2013}.)

\subsection{\label{subsec:Energy_gap}Band gap}

%=============================================================================================================
\begin{table}
\caption{\label{tab:table5} Energy difference (in meV) between the considered extrema. Asterisks mean absence of the X extremum (and additionally the B extremum in the GGA(0)+$GW$ case) in the corresponding calculations done in this work. The fundamental band gap is set in bold. }
\begin{ruledtabular}
\begin{tabular}{lcccc}
            Calculations   &$\Delta_{B-C}$&$\Delta_{B-X}$&$\Delta_{A-C}$&$\Delta_{A-X}$  \\
  \hline
  GGA(0)         &    \textbf{65}        &   $\ast$     &     73       &     $\ast$     \\
  GGA(0)+$GW$      &  $\ast$      &   $\ast$     &     \textbf{76}       &     $\ast$     \\
  GGA(II)        &   \textbf{114}        &    126       &    142       &      154       \\
  GGA(II)+$GW$     &   168        &    243       &    \textbf{156}       &      231       \\
  LDA(0)         &    \textbf{41}        &   $\ast$     &     79       &     $\ast$     \\
  LDA(I)         &   \textbf{99}        &    \textbf{99}       &    135       &      135       \\
  LDA(I)+$GW$      &   \textbf{109}        &   $\ast$     &    112       &     $\ast$     \\
  LDA(II)        &   100        &     \textbf{79}       &    146       &      125       \\
  LDA(II)+$GW$     &   \textbf{102}        &   $\ast$     &    123       &     $\ast$     \\
 % Experiment     &              &              &              &                \\
  FLAPW sX-LDA\footnotemark[1]  &           &   \textbf{154}     &            &           \\
  PWP LDA\footnotemark[2]       &           &   \textbf{87}     &            &           \\
  PWP LDA+$GW$\footnotemark[2]  &   \textbf{165}        &   166     &            &          \\
\end{tabular}
\footnotetext[1]{From Ref.~\onlinecite{Kim_Freeman_PRB_2005}}
\footnotetext[2]{From Ref.~\onlinecite{Kioupakis_PRB_2010}}
\end{ruledtabular}
\end{table}
%=============================================================================================================

Now we additionally discuss the band gap as obtained for bismuth telluride with taking into account the $GW$ corrections to the GGA and the LDA band structure. First, it is worth emphasizing that in all the $GW$ calculations we found the extrema B and C to be located very close to each other in the mirror plane. These extrema with $M=6$ form a direct band gap that is the largest in the GGA(II)+$GW$ case (see Table \ref{tab:table5}) and, excepting this very case, is a fundamental gap. As compared with the respective DFT calculations, the $GW$ corrections enlarge the fundamental band gap of the considered topological insulator, i.e., in that sense act as in the case of the conventional semiconductors. The LDA(I)+$GW$ calculations yield the largest \textit{direct fundamental gap}, while the GGA(II)+$GW$ results demonstrate the largest \textit{indirect fundamental gap}. The latter is close to those in Ref.~\onlinecite{Kim_Freeman_PRB_2005} with the only difference that our indirect gap is formed by the extrema A ($M=2$) and C ($M=6$), while in Ref.~ \onlinecite{Kim_Freeman_PRB_2005} both extrema have the multiplicity of 6.

In all the calculations performed, due to the $GW$ corrections the extremum C becomes more prominent independently of the relative position of the latter and the extremum X on the DFT level. Note that such an effect (with a smaller strength) is traced in Ref.~\onlinecite{Kioupakis_PRB_2010} too. This preserves the multiplicity of the VBM, but makes unlikely a formation of an \textit{indirect fundamental} band gap with the CBM with $M=6$, as it comes from the Shubnikov-de Haas measurements. Nevertheless, among all the presented results, we would like to mark out our GGA(II)+$GW$ calculations which provide us with the band gap comparable with the experimental values and with the profiles of the CB and the VB similar to those appearing in ARPES measurements of Ref.~\onlinecite{Chen_Science_2009} (see also the respective discussion in Sec.~\ref{sec:introduction}). Note that, as follows form our study, the relative position of the CB extrema on energy scale can be affected even by mild temperature effect, which can change the band-gap character (from indirect to direct) revealed in the GGA(II)+$GW$ case.

\section{Conclusion}

In conclusion, on the equal footing within density functional theory we analyzed how the atomic positions and the approximation chosen for the XC functional affect bulk band-gap values, locations of valence- and conduction-band extrema in the Brillouin zone, and dispersion of these bands in the vicinity of the extrema in Bi$_2$Te$_3$. We showed that at fixed atomic positions the LDA yields energy differences between the VB and CB extrema, which are systematically smaller than the GGA does. For a given approximation for the XC functional, alterations of the atomic positions upon relaxing at fixed unit-cell volume lead to increasing these differences. Such an increase is accompanied by a steeper dispersion along the $\Gamma$-Z line and by enlarging the band-inversion region in the vicinity of the $\Gamma$ point. Upon relaxing the atomic positions, the number of the valence-band extrema becomes greater owing to the appearance of a quite extensive VB maximum labeled by X. This maximum is a local one in the GGA calculations, whereas in the LDA calculations it plays a role of the VBM. Both the GGA and the LDA results predict the multiplicity of the VBM and the CBM in agreement with the Shubnikov-de Haas measurements. The locations of the extrema in the mirror plane of the BZ undergo slight changes, whether we move from the GGA to the LDA or use different sets of the atomic positions. On the DFT level, we thus demonstrated that the examined characteristics of the electronic structure of Bi$_2$Te$_3$ are quite delicate to study with \textit{ab-initio} methods.

For each set of the atomic positions considered on the DFT level and with different exchange-correlation functionals, we calculated many-body corrections within the one-shot $GW$ approach. We showed that at fixed atomic positions the one-shot $GW$ results depend on the DFT reference one-particle band structure. This dependence is more strongly than in the case of bismuth selenide,\cite{Nechaev_PRBR_2013} what may stimulate a further study of bismuth telluride but already beyond the one-shot perturbative approach. We found that for Bi$_2$Te$_3$ the one-shot $GW$ corrections enlarge the fundamental band gap and bring its value in close agreement with experiment in the case of the relaxed atomic positions and GGA reference one-particle band structure.

We have noticed that only the use of the relaxed atomic positions gives adequate $GW$ results. For those cases, we demonstrated that in Bi$_2$Te$_3$ as in the conventional semiconductors the band-gap enlargement is mainly caused by ``moving'' the conduction band away from the valence band on the energy scale. Due to the band inversion, the movement apart causes a $\Gamma$-band-gap reduction without inducing a crucial rearrangement of VB extrema as it occurs in, e.g., Bi$_2$Se$_3$, where the VBM shifts from a location in the mirror plane and far from the BZ center to $\Gamma$. As a consequence, we can infer that, in general, on the DFT level the dispersion of the band-gap edges in bismuth telluride can be described adequately. This means that, first, in the case of Bi$_2$Te$_3$ there is no strong reason to call for a revision of experimental results which were interpreted on the basis of a DFT study. Second, as distinct from the conventional semiconductors, for the three-dimensional topological insulators as a class of materials one cannot \textit{a priori} say to what effects the $GW$ corrections may lead.

We have revealed that with the $GW$ corrections in the relaxed cases the mentioned extremum X that appears on the DFT level becomes less evident in the GGA-based calculations or disappears if the LDA reference one-particle band structure is used. In all the $GW$ calculations, the valence band is represented by the maximum with $M=6$, which in the mirror plane is located practically at the same point as the conduction-band extremum that is the CBM in the LDA-based calculations. In the GGA-based calculations with the relaxed atomic positions, this extremum is merely 12 meV higher than the global CB minimum located on the $\Gamma$-Z line.

A comparison of the effective-mass tensor parameters calculated for the found extrema with the experimental ones revealed that in general the $GW$ corrections bring their in better agreement (especially, in the case of the in-plane components of the tensor) if the GGA reference one-particle band structure is used. However, the best agreement is reached for the extremum X that is not the VBM in our $GW$ calculations. The effective-mass results have also demonstrated that the extremum lying on the $\Gamma$-Z line cannot correspond to the CBM as recognized from the Shubnikov-de Haas measurements. Nevertheless, we stated that on the strength of all the considered electronic-structure characteristics our GGA-based $GW$ calculations performed for the relaxed atomic positions give the most adequate picture of the valence- and conduction-band profile (including the energy gap between these bands) in the $\bar{\Gamma}-\bar{M}$ direction of the two-dimensional Brillouin zone, which resembles that coming from available ARPES measurements.

\section*{\label{sec:acknowledgments}Acknowledgments}

We thank I. Aguilera, C. Friedrich, and S. Bl\"{u}gel for a critical reading of the manuscript and helpful discussions. We also acknowledge partial support from the Basque Country Government, Departamento de Educaci\'{o}n, Universidades e Investigaci\'{o}n (Grant No. IT-366-07), and the Spanish Ministerio de Ciencia e Innovaci\'{o}n (Grant No. FIS2010-19609-C02-00), and the Ministry of Education and Science of Russian Federation (Grant No. 2.8575.2013).

%=============================================================================================================
%=============================================================================================================
%=============================================================================================================


\begin{thebibliography}{00}
\bibitem{remark_coords} We imply the real-lattice vectors $\mathbf{R}_1=(0,\,-a_H/\sqrt{3},\,c_H/3)$, $\mathbf{R}_2=(a_H/2,\,a_H/(2\sqrt{3}),\,c_H/3)$, $\mathbf{R}_2=(-a_H/2,\,a_H/(2\sqrt{3}),\,c_H/3)$, and the reciprocal-lattice vectors $\mathbf{K}_1=2\pi(0,\,-2/(\sqrt{3}a_H),\,1/c_H)$, $\mathbf{K}_2=2\pi(1/a_H,\,1/(\sqrt{3}a_H),\,1/c_H)$, $\mathbf{K}_3=2\pi(-1/a_H,\,1/(\sqrt{3}a_H),\,1/c_H)$, where the lattice parameters of the hexagonal unit cell $a_H$ and $c_H$ are related to the lattice parameters of the rhombohedral unit cell $a_{Rh}$ and $\vartheta_{Rh}$ as $a_H=2a_{Rh}\sin(\vartheta_{Rh}/2)$ and $c_H=a_{Rh}\sqrt{3(1+2\cos\vartheta_{Rh})}$.
\end{thebibliography}
\end{document}